\documentclass[preprint,twocolumn]{aastex61}
\usepackage{amsmath,amssymb,amsthm}
\usepackage{graphicx}
\usepackage{CJKutf8}

\newcommand{\depletiontime}{T_{\rm dep}}
\newcommand{\dampingtime}{T_{\rm damp}}

\newcommand{\mjupiter}{M_{\rm J}}
\newcommand{\msaturn}{M_{\rm S}}

\shortauthors{Zheng, Lin, Kouwenhoven, Mao \& Zhang}

\begin{document}
\begin{CJK*}{UTF8}{gbsn}

\title{Clearing residual planetesimals by sweeping secular resonances
in transitional disks: a lone-planet scenario for the wide gaps in 
debris disks around Vega and Fomalhaut}

\author{Xiaochen Zheng (郑晓晨)}
\affiliation{Department of Physics and Center for Astrophysics, Tsinghua University, Beijing 10086, P.R. China}
\email{x.c.zheng1989@gmail.com}
\author{Douglas N. C. Lin (林潮)}
\affiliation{Institute for Advanced Studies, Tsinghua University, Beijing 10086, P.R. China}
\affiliation{Department of Astronomy and Astrophysics, University of California, Santa Cruz, CA 95064, USA}
\affiliation{National Astronomical Observatories of China, Chinese Academy of Sciences, 20A Datun Road, Beijing 100012, P.R. China}
\author{M. B. N. Kouwenhoven (柯文采)}
\affiliation{Department of Mathematical Sciences, Xi'an Jiaotong-Liverpool University, 111 Ren'ai Rd., Suzhou Dushu Lake Science and Education Innovation District, Suzhou Industrial Park, Suzhou 215123, P.R. China}
\author{Shude Mao (毛淑德)}
\affiliation{Department of Physics and Center for Astrophysics, Tsinghua University, Beijing 10086, P.R. China}
\affiliation{National Astronomical Observatories of China, Chinese Academy of Sciences, 20A Datun Road, Beijing 100012, P.R. China}
\affiliation{Jodrell Bank Centre for Astrophysics, School of Physics and Astronomy, The University of Manchester, Oxford Road, Manchester M13 9PL, UK}
\author{Xiaojia Zhang (张晓佳)}
\affiliation{National Astronomical Observatories of China, Chinese Academy of Sciences, 20A Datun Road, Beijing 100012, P.R. China}

\begin{abstract}
Extended gaps in the debris disks of both Vega and Fomalhaut have been 
observed.  These structures have been attributed to tidal perturbations 
by multiple super-Jupiter gas giant planets.  Within the current 
observational limits, however, no such massive planets have been detected. Here 
we propose a less stringent `lone-planet' scenario to account for the 
observed structure with a single eccentric gas giant and suggest that 
clearing of these wide gaps is induced by its sweeping secular resonance.  
With a series of numerical simulations, we show that the gravitational 
potential of the natal disk induces the planet to precess.  At the 
locations where its precession frequency matches the precession frequency
the planet imposes on the residual planetesimals, their eccentricity is 
excited by its resonant perturbation. Due to the hydrodynamic drag by 
the residual disk gas, the planetesimals undergo orbital decay as their 
excited eccentricities are effectively damped. During the depletion of 
the disk gas, the planet's secular resonance propagates inward and clears 
a wide gap over an extended region of the disk.  Although some residual 
intermediate-size planetesimals may remain in the gap, their surface 
density is too low to either produce super-Earths or lead to 
sufficiently frequent disruptive collisions to generate any observable dusty 
signatures.  The main advantage of this lone-planet sweeping-secular-resonance 
model over the previous multiple gas giant tidal truncation scenario 
is the relaxed requirement on the number of gas giants. The observationally 
inferred upper mass limit can also be satisfied provided the hypothetical 
planet has a significant eccentricity.  A significant fraction of solar
or more massive stars bear gas giant planets with significant eccentricities.
If these planets acquired their present-day kinematic properties prior to 
the depletion of their natal disks, their sweeping secular resonance 
would effectively impede the retention of neighboring planets and 
planetesimals over a wide range of orbital semi-major axes.
\end{abstract}



\section{Introduction}

Debris disks are common around nearby young stars \citep{morales2013}. 
Infrared and sub-millimeter observations have revealed that some of these
disks have wide ($> 10$~AU) gaps separating a compact (a few AU in size) 
inner disk from an extended outer ring.  For example, the observed excess 
emission for wavelengths $\lambda \ge 15\ \mu$m in the debris disk 
around $\epsilon$~Eri indicates the presence of a gap between two 
narrow rings \citep{backman2009}. The debris disk around HR8799 contains
an inner warm belt, a broad cold belt and an outer halo \citep{su2009}.
Two additional debris systems with widely separated inner disks and outer 
rings were found around Vega and Fomalhaut \citep{su2013}. Just like  
HR8799, the central stars of these disks are all more massive than the 
Sun.  More recently, \cite{morales2013} announced the discovery of 
four additional debris disks with widely separated belts around HD70313, 
HD71722, HD159492 and HD104860. 

These two-component debris disks show some resemblance to the kinematic 
architecture of our own Solar system, which contains the inner asteroid belt and the outer Kuiper belt. Between these two
belts, gravitational perturbation of the two gas giants and two ice giants
may have destabilized the orbits and cleared residual planetesimals in the 
region both during their formation epoch and during the subsequent dynamical evolution of the Solar System \citep{duncan1989}. In particular, after the solar nebula was removed, large amount of material was swept inward producing a high radial mass concentration in the inner region of the Solar system, finally leading to the formation of terrestrial planets and leaving small amount of debris known as the asteroid belt we observe today \citep{thommes2008, zheng2017, bromley2017}. The dynamical sculpting process of the multiple giant planets on the formation of the present-day's Kuiper Belt, may have proceeded over several $10^8$ yrs \citep{levison2008}.

This similarity suggests that it may be possible to infer the presence 
of one or more exoplanets from the surface brightness distribution of 
disks around other stars.  Indeed, \cite{kalas2008} reported a possible 
faint planetary companion around Fomalhaut.  It has been attributed as 
a potential culprit for dynamically shaping the ring-like outer debris 
disk around Fomalhaut \citep{chiang2009}.  But, this interpretation on 
the physical nature of this candidate remains controversial 
\citep{janson2012, boley2012}.  Regardless its validity, its mass is too
low to account for the wide gap between the inner and outer regions of Fomalhaut's disk. 

In order for single gas giants to open a wide gap in the debris disks 
around Vega and Fomalhaut, their masses must exceed several tens of Jupiter masses. 
Such conspicuous massive companions have not been found \citep{marois2006, 
heinze2008, su2013}.  Partial clearing of planetesimals and grains
over an extended region may also be induced by single less massive 
(still many times that of Jupiter) planets on highly eccentric ($e > 0.8$)
orbits.  The excitation, stability, and retention of such massive, 
highly eccentric, single planets with such large eccentricities 
and semi-major axis remain an outstanding issue. 

\cite{su2013} attributed the combined tidal perturbation of one or 
multiple planets as the culprit for the large dust-free gap in 
the debris disks.  Each planet has a chaotic zone within which
the orbits of other planets and planetesimals are destablized by long-term
dynamical instabilities.  Similar to the Solar System, four or more 
Jupiter-mass planets are needed to fill the wide gap with their 
overlapping chaotic zones.  But, the analysis of the combined data 
from direct imaging, microlensing, and radial velocity survey shows 
a precipitous decline in the abundance of planets with masses 
larger than Neptune in all period ranges \citep{clanton2016},
although there are a few known gas giants beyond $\sim 100$~AU from the 
host stars. In contrast, Neptune-mass/size planets are common at 
a distance of less than 1AU from their host stars. A slightly 
sparser population of distant Neptune-mass planets has been 
extrapolated from the statistical properties of their close-in 
cousins. This inference has not been observationally verified 
because these planets have masses which are below the detection 
threshold. Since the width of the chaotic zone around any host 
stars increases with their masses, additional Neptune-mass planets
(in comparison with Jupiter-mass planets) are needed to provide 
an adequate filling factor of their chaotic zones over the 
wide gap regions.

The key challenge for clearing wide gaps by embedded planets is 
their limited zone of influence. Due to secular interactions, eccentric
planets can perturb other planets and residual planetesimals beyond 
the chaotic zone.  These perturbations are generally weak.  However,
gravitational interactions with the natal disks also leads to the precession of 
embedded planets. The planets' mutual secular perturbations also 
induce nodal precession and eccentricity modulation.  
In resonant locations where the disk-induced precession rate
matches to that due to planets' interaction with each other, 
persistence of the relative longitude of periapses of the
interacting planets can lead to large eccentricities.

Through secular interactions, multiple planets undergo eccentricity
modulations and apsidal precession with distinct eigen frequencies
\citep{murray1999}. In the solar system, Jupiter and Saturn 
also impose secular perturbations to asteroids in the main belt.  In
some special locations, asteroids precess at rates close to gas giants' 
apsidal precession rate.  These asteroids attain longitudes of 
periapse at a nearly constant (non-zero) phase relative to that of 
Jupiter such that their eccentricities are largely excited.  Although 
these zones of ($\nu_5$ and $\nu_6$) secular resonances are narrowly 
confined, they can extend well outside the chaotic zones.  
Furthermore, the location of the secular resonances can evolve with  
the eigen-frequencies of the system.  The Nice model is constructed 
based on the assumption that, due to a hypothetical widening of 
Jupiter-Saturn separation, the $\nu_5$ secular resonance may have 
propagated between the main belt region and its current location 
(interior to the orbit of Venus). \cite{gomes2005} proposed that, 
along the propagation paths of the $\nu_5$ secular resonance, 
the eccentricities of some asteroids were greatly excited so that 
they were cleared out of the main belt region and became the
culprits of the Lunar ``late heavy bombardment''.  This hypothesis 
remains controversial because it is also likely to excite the 
eccentricities of the terrestrial planets to values much higher
than the present-day observed values \citep{brasser2009, agnorlin2012}.

There are other effects which can lead to the efficient clearing
of planetesimals and planets.  Gas giant planets were formed in 
gas-rich protostellar disks. With sufficient masses, the tidal 
perturbation of these embedded gas giants induce the formation 
of a gap in the disk's gas distribution \citep{bryden2000}. 
Hydrodynamic simulations show that this process also leads to dust 
clearing and filtration.  In the case of single-planet systems, 
the width of the gap is typically a few Hill radii, and less than 
the distance between the planet and its 2:1 mean motion resonance 
\citep{zhu2011, zhu2012}.  With this limited perturbation on the 
gas and dust distribution, many hypothetical gas giants would be 
needed to clear the wide gaps in the disks around Vega and Fomalhaut.  

The gravity of the planets' natal disks can also lead to the precession
of their orbits.  In a minimum mass nebula this effect dominates the
secular perturbation between planets. As the disks' contribution to
the total gravity reduces during the depletion, the planets' 
precession rate due to the disk potential declines.  Consequently, 
the location of their secular resonance propagates over wide regions.
In the Solar System, the $\nu_5$ and $\nu_6$ secular resonances 
sweep across the inner regions of the solar nebula (where the 
terrestrial planets are located). This effect leads to the excitation 
of residual planetesimals' eccentricities by the major planets, 
even at those large distances \citep{heppenheimer1980, ward1981, 
nagasawa2003}. 

The diminishing residual disk gas also damps the eccentricity of the 
planetesimals, which leads to the decay of their orbits \citep{nagasawa2003, 
nagasawa2005}. Since the $\nu_5$ and $\nu_6$ secular resonances interior
to the gas giants' orbits also propagate inwards, the residual 
planetesimals caught along their paths continue to migrate inward.  This 
process may have led to (i) a significant clearing of the 
asteroid belt \citep{zheng2017}; (ii) a high concentration and dynamical 
shake-up of planetesimals interior to the orbit of Mars \citep{zheng2017}; 
and (iii) it may have promoted the formation and subsequent eccentricity damping 
of the terrestrial planets \citep{thommes2008}.  This model is consistent 
with the formation time scale of the terrestrial planets ($\sim 10^{7-8}$ 
yr) inferred from radioactive isotopes and the small orbital eccentricities 
($e < 0.1$) of the terrestrial planets.

Here, we propose that the observed wide gaps in some debris disks may have been 
cleared by the secular resonances of a known planetary candidate (in 
Fomalhaut) or yet to be detected (around Vega and other systems) 
isolated or multiple gas giants or super Neptunes, as the secular resonances 
sweep across this region during the depletion of the gas disk. Analogous 
to the asteroids, excited eccentricities of the residual planetesimals 
in the region between the inner disk and outer ring are mainly damped 
by the hydrodynamic drag of the diminishing disk gas.  In addition 
to the extended gap, this scenario offers a potentially exciting prospect 
of finding signatures of ongoing planetary assemblage in debris disks 
around stars with an age comparable to the formation epoch of 
terrestrial planets in the solar system \citep{su2009}.

Since not a single planet with a mass in excess of $\sim 1 M_J$ (an 
observationally inferred upper limit) has been found around Vega
and Fomalhaut, we explore here the possibility that the extensive 
gaps are produced by a sweeping secular resonance of a lone-planet
with a modest mass and eccentricity in the debris disks around these 
and other similar systems. Although systems of multiple Neptune-mass
planets may also lead to a similar effect, we opt the lone-planet
scenario as an idealized simplest possibility.

This paper is organized as follows: 
in \S\ref{section:vega}, we briefly introduce the lone-planet 
possibility in the Vega system, including detailed physical models 
to mimic the dust-free gap opening process by lone-planet scenario 
in Vega system, develop a rough method to constrain planet 
candidates based on observed features, and provide predictions 
for future planet detection. In \S\ref{sec:fomalhaut}, 
we further extend our model to Vega-like system, Fomalhaut, 
inferring the properties of the prospective planet and compare 
these with a planet candidate, Fomalhaut b. A summary of our
results and an extensive discussion on the other two belt debris 
disks are presented in \S~\ref{section:conclusions}.


\section{A Lone-planet scenario for the Vega system} \label{section:vega}

\subsection{Dynamical causes for wide dust-free gaps in debris disks}

\begin{figure*}
\centering
\includegraphics[width=1\textwidth,clip=true]{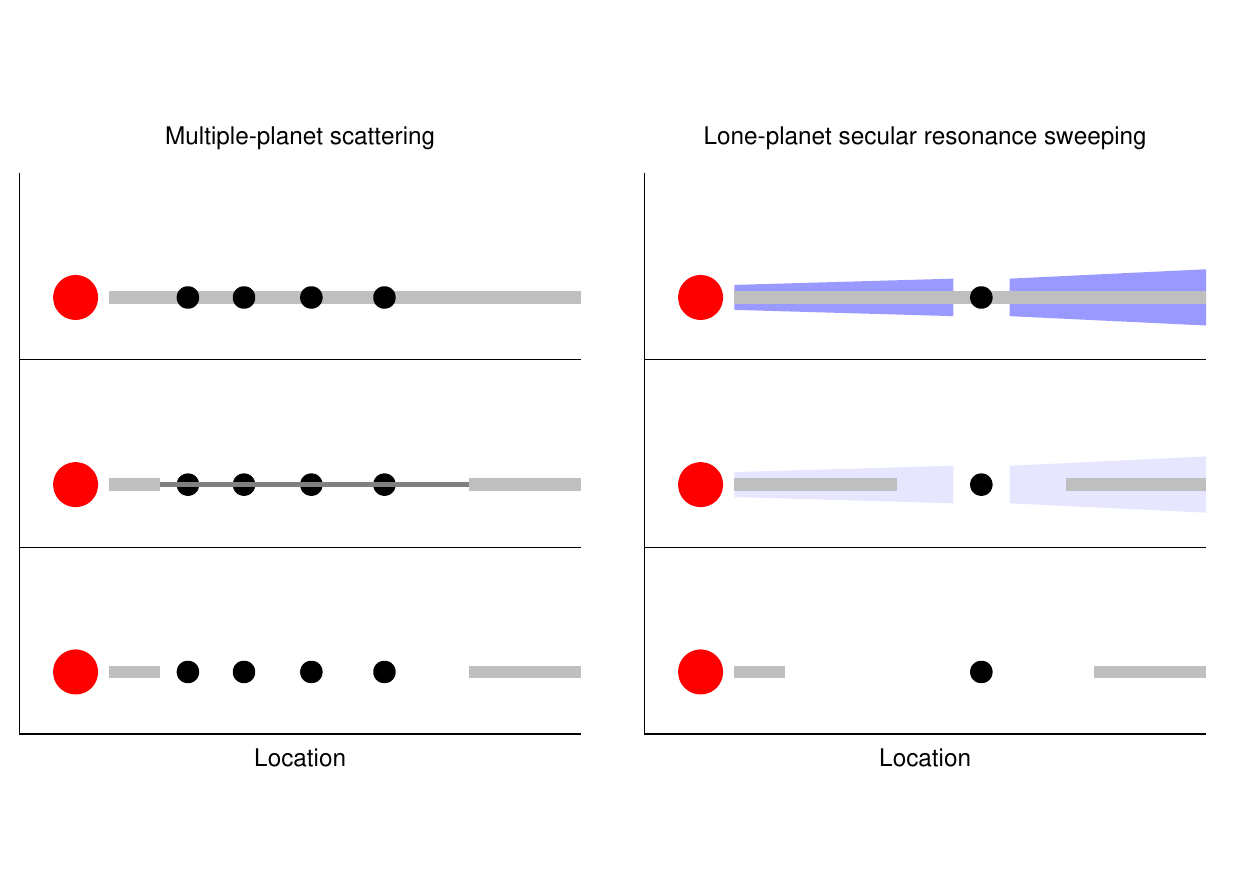}
\caption{A schematic representation comparing the multiple-planet
scenario (left) with the lone-planet scenario (right). The red 
solid dots refer to the host star, while gray bars and blue-shaded 
areas label the possible distribution of planetesimal disk and a 
depleting gas disk, respectively. From top to bottom, two models 
both describe the orbital evolution of a planetary system from an 
early age to the present day. The thinning (or shrinking) of the 
gray bars on planetesimal disk as well as the fading blue areas 
of the gas disk indicate their decreasing surface density.}
\label{fig:two_scenario}
\end{figure*}

Infrared observations indicate that Vega's debris disk contains a 
large gap that spans from $\sim 15$~AU to $\sim 110$~AU \citep{su2013}. 
The relatively large size of the gap in Vega's debris disk can be used 
to constrain the orbital configuration of Vega's (hypothetical) 
planetary system. In our solar system, the four gas giants cleared 
most of the debris between the outer asteroid belt $\sim 3$~AU and 
the inner Kuiper belt $\sim 35$~AU. Similarly, in the young HR 8799 
system, (at least) four giant planets have been detected 
\citep{marois2008, marois2010} and they are believed to be the chief 
culprit for the large separation between the warm ($\sim 6 - 15$~AU) 
and cold ($\sim 90 - 300$~AU) debris disks \citep{su2009}. Therefore, 
a multiple-planet configuration is a preferred scenario in the 
Vega and Vega-like systems which has a characteristic debris disk 
with two broadly separated components. \cite{su2013} claims that 
the deficit of planetesimals orbiting Vega with semi-major axes 
in the range $15-110$~AU can be attributed to the existence of 
multiple, currently undetected planets. However, in our study 
we discuss the possibility of a singe planet with an intermediate 
mass (few $\mjupiter$) and intermediate orbital eccentricity that 
can produce this gap. 

Motivated by earlier works on the dynamic shake-up model 
\citep{nagasawa2005, thommes2008}, we focus on the dynamical 
evolution of a debris disk around the star, in the presence 
of a single gas giant and a depleting circumstellar gas disk. 
The evolution of the debris disk (planetesimal disk) is affected 
by the gravitational potential of both the gas giant and the gas 
disk, and also by the hydrodynamical drag due to the protoplanetary 
gas disk. Generally, secular resonances occur when the precession 
rate of the gas giant caused by the gas disk potential matches that 
of planetesimals in the debris disk, whose precessions are 
modulated by both the gas disk and the gas giant. After being 
captured into secular resonance, the planetesimals are excited 
and generally migrate away from their original orbits due to 
subsequent damping. As the gas disk depletes over  time, the 
location of the secular resonance also sweeps inward, through 
the debris disk. This provides a possible explanation for 
the large dust-free gap structures in the debris disks of 
Vega-like systems. In order to clear the gap region, it is 
therefore necessary for the secular resonance to sweep through
the entire region where the dust infrared emission is observed 
to be absent.

Even though the evolution process especially the planetesimal 
clearing mechanisms of these two scenarios are quite different, 
theoretically, both models can reproduce the present-day debris 
disk observation (see Figure~\ref{fig:two_scenario}).

\subsection{Model Setup}

\subsubsection{The Circumstellar Gas Disk}

As we aim at the dynamical evolution of a swarm of planetesimals 
which are embedded in a depleting gas nebula, we modified the 
publicly available HERMIT4 package \citep{aarseth2003} by adding 
an analytical disk potential and hydrodynamic drag on particles, 
as described in \cite{zheng2017} for details. Briefly, a radial 
surface density distribution for the gas disk of the form 
\begin{equation} 
\Sigma(r, t) = \Sigma_{0} \, \exp(-t/\depletiontime)  \, r^{-k} \ ,
\label{eq:sigma}
\end{equation}
is adopted, where $\Sigma_0$ is the fiducial surface density of 
gas disk at 1~AU, and $\depletiontime$ is the depletion time scale 
of the gas disk. In the \emph{minimum mass nebula model} 
\citep{hayashi1985}, $\Sigma_0 = 1,000 - 2,000\ \rm{g\,cm^{-2}}$ 
and a power law index $k = 1.5$ are widely used for describing 
our solar system and solar-like systems at early times 
\citep[e.g.,][]{nagasawa2005}. The best estimate for Vega's
age is around 450 Myr \citep{yoon2010}. A representative 
depletion timescale, $\depletiontime = 1-5~$Myr, is mainly 
discussed in this paper.

The presence of a ($<$ few) Jupiter-mass planet not only perturbs 
nearby planetesimal disk, but its tidal interaction with gas also 
affects the morphology of the disk by opening a gap in the 
surface density distribution \citep{linpapaloizou1986}. The gap 
structures (the depth and width of the gap) are closely related 
to the planet's eccentricity and mass, especially for a planet 
with an eccentricity above its Hill radius,  
$r_{\rm Hill} \sim a_{\rm p} ({\frac{m_{\rm p}}{M}})^{\frac{1}{3}}$, 
divided by its semi-major axis $a_{\rm p}$ \citep{hosseinbor2007}. 
For computational simplicity, we conservatively estimate a gas-free gap 
in the region around $a_{\rm p} \pm r_{\rm Hill}$.

\subsubsection{The Planetesimal Disk}

In our model, a planetesimal disk (debris disk) coplanar to the assumed planet and overlaps with the gas disk. We assume that all planetesimals and gas particles within the Hill radii region of the hypothetical planet are totally accreted (or depleted) , the motion of planetesimals is determined by the central star and the disk's gravity beyond the gap region.  Since most planetesimals are located outside the 
gap region, they are subject to the disk's self gravity from nearby regions
\citep{nagasawa2005,zheng2017}. The eccentricity of planetesimals is also 
damped by their tidal interaction with the residual disk gas. When the total 
damping timescale, $\dampingtime$, is shorter than or comparable to the gas 
depletion timescale, $\depletiontime$, of the protostellar disk, the 
planetesimals migrate inwards with the secular resonances' inwardly 
sweeping rate.  This combined influence of the planet's secular perturbation
and the disk gas' eccentricity damping clear planetesimals over large
region of the disk.

Throughout this work, we consider km-size planetesimals.  These particles
are the most likely parent bodies of $\mu m$ and mm-size collisional 
fragments which are mostly to be responsible for the observed inferred
features of the Vega and Vega-like system \citep{wyatt2002}. In this size 
range, eccentricity damping of planetesimals is mainly caused by the 
hydrodynamic drag, which is inversely proportional to the planetesimal 
radius. Therefore, if the hypothetical planet can efficiently clear
the $\sim$~km planetesimals, it would also be effective in opening up a 
wide gap for the sub-km size planetesimals.

\subsection{Boundary Constraints}

Under the assumption that there is indeed a single, undetected giant planet in the debris disk, 
the observed disk boundaries provide constraints on its orbital parameters (in particular, its mass $m_{\rm p}$, semi-major 
axis $a_{\rm p}$ and eccentricity $e_{\rm p}$). In general, the presence of a gas giant planet can generate a chaotic zone 
surrounding its orbit. Within this zone, it is essentially impossible 
for planetesimals to attain stable orbits. The dynamical origin for this 
chaotic zone is mainly due to the overlap of mean motion resonances 
\citep{wisdom1980}.  The width of the chaotic zone is related to the 
planet-to-star mass ratio, $\mu$. According to \cite{morrison2015}, 
particles in a planet's chaotic zone are mostly ($95\%$) driven out.
For an eccentric planet with $e_{\rm p} > 0.2$ and $ 10^{-9} \le 
\mu \le 10^{-1.5}$, the interior and exterior boundaries for the 
chaotic zone are not asymmetric about the planet's orbit and they
are located at separations
\begin{equation}
\delta a_{\rm int} \approx 1.2 \, \mu^{0.28} \, r_p ,
\label{eq:a_int}
\end{equation}
and 
\begin{equation}
\delta a_{\rm ext} \approx 1.7 \, \mu^{0.31} \,r_a ,
\label{eq:a_ext}
\end{equation}
respectively, where $r_p = a_{\rm p} (1 - e_{\rm p})$ and
$r_a = a_{\rm p} (1 + e_{\rm p})$ are its periastron and apoastron
distance from its host star.  \cite{su2015} have suggested that
the wide gap between the outer boundary of the warm and inner boundary 
of the cold debris belt around HD~95086 may due to the clearing
of planetesimals in the chaotic zones of multiple hypothetical planets in 
addition to the confirmed exoplanet \citep{rameau2013, rameau2016}.
Similar scenario of dynamical clearing by multiple, massive, undetected 
planets have been invoked to account for the wide, dust-free gaps in several 
other systems including the $\epsilon$ Eridani system \citep{su2017} and 
the HIP 67497 system \citep{bonnefoy2017}. 

In the lone-planet scenario, we suggest that residual planetesimals 
may be cleared well beyond the region between $a_{\rm int}$ and 
$a_{\rm ext}$ during the epoch of disk depletion. The disk's 
gravitational potential leads to precession of any embedded planet's
orbit.  As this contribution to the gravity weakens with the decline in
the disk's surface density, the planet's precession slows down and the 
location of its secular resonance expands from the proximity of its 
orbit to far-flung regions in the disk.  During the passage of the
secular resonance, the planetesimals' eccentricity is excited as their
angular momentum is removed by the planet's tidal torque.  Subsequent
damping of the planetesimals' eccentricities by the hydrodynamic drag of the
residual disk gas dissipates their orbital energy and induces them to
undergo orbital decay.   

\begin{figure*}[!]
\centering
    \includegraphics[width=1\textwidth]{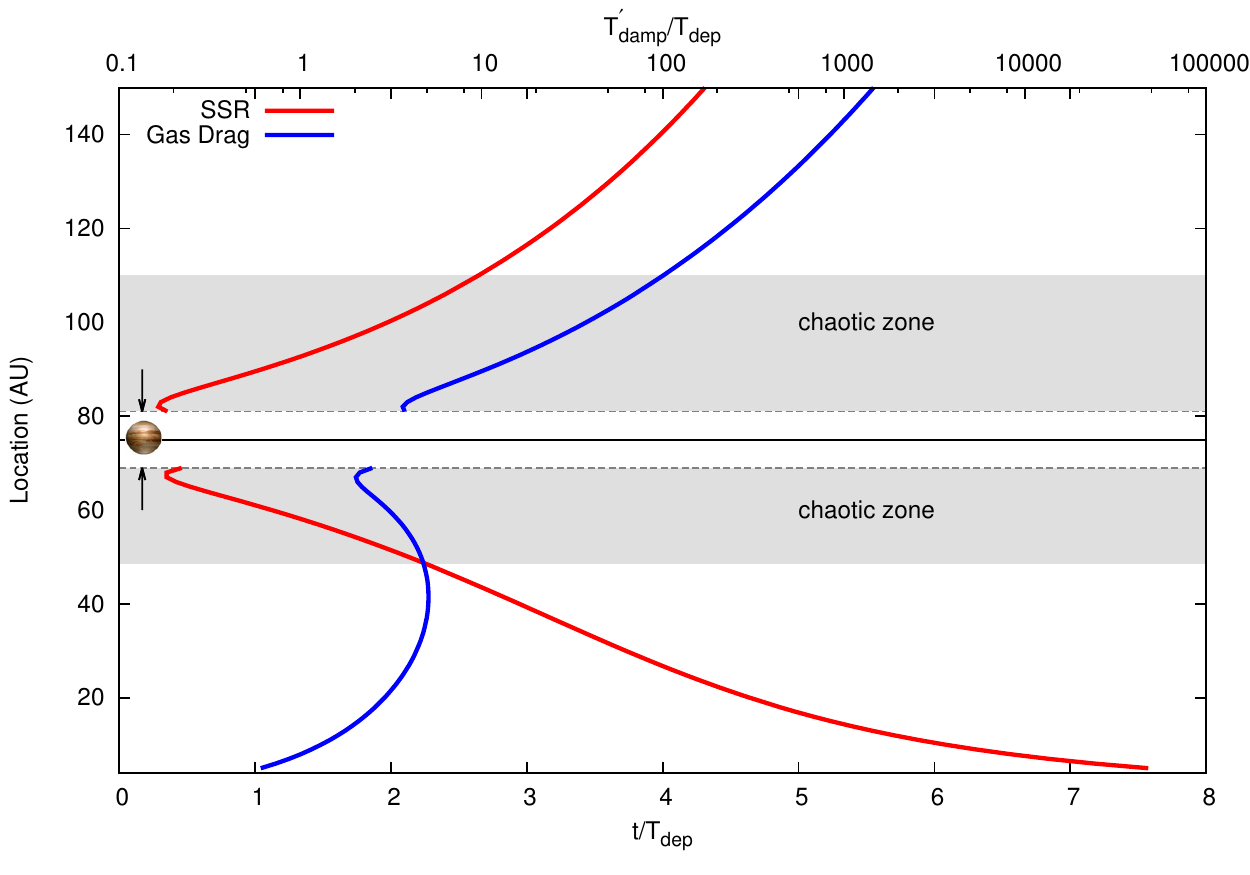}  \\
\caption{Location of a hypothetical planet's inner and outer secular
resonances (red curve) as a function of the depletion factor
$t/\depletiontime$.  The depletion timescale $\depletiontime=5$ Myr,
the planet's $m_{\rm p}=3 M_{\rm J}$, $a_{\rm p} = 75$ AU, and 
$e_{\rm p}=0.2$ so that its $a_{\rm int} \sim 50$ AU and $a_{\rm ext}
\sim 110$ AU.  The location of secular resonances have propagated beyond
$a_{\rm int}$ and $a_{\rm ext}$ at $t/\depletiontime \simeq 3$.
At the location of the secular resonance, the eccentricity damping 
efficiency is inversely proportional to the magnitude of 
$\dampingtime/\depletiontime$ (blue curve).  The solid line
represent the orbital semi major axis of the planet and the 
dotted black lines indicate the width of a gas-free zone. 
}
\label{fig:ssrlocation}
\end{figure*}

In the region interior to the planet's orbit, the inwardly migrating
planetesimals endure a prolonged resonant perturbation as their orbital 
evolution proceeds in the same direction as the propagation of the 
planet's inner secular resonances. Despite the gas depletion, an efficient 
damping rate of planetesimals' eccentricity is maintained as they enter 
into the dense inner regions where $\dampingtime / \depletiontime < 10$  
 (for example, see Figure \ref{fig:ssrlocation}).  Consequently, residual 
planetesimals interior to the planet's orbit can migrate over extensive 
distances and the inner boundary of the dust-free gap is strongly 
affected by the sweeping secular resonance mechanism.

In the region outside the planet's orbit, the outward propagation of its
secular resonance diverges from the orbital decay of the planetesimals
along its path. During the single passage of the secular resonance, the 
amplitude of eccentricity excitation of the perturbed planetesimals is 
relatively modest (see also Figure \ref{fig:ssrlocation}).  
As the secular resonance expands to large radii, the local 
eccentricity damping rate of the perturbed planetesimals also diminishes 
with the decrease in the surface density of the disk gas such that 
$\dampingtime / \depletiontime \gg 1$ (see Eq.~\ref{eq:sigma}). The 
weakened contribution of the sweeping 
secular resonances implies that the outer boundary of the cleared 
region may be mainly determined by the stability condition in Equation
(\ref{eq:a_ext}), although some planetesimals may diffuse into the chaotic
zone shortly after the passage of the secular resonance.  

\begin{figure*}[!]
\centering
  \begin{tabular}{p{0.5\textwidth}p{0.5\textwidth}}
    \includegraphics[width=0.5\textwidth]{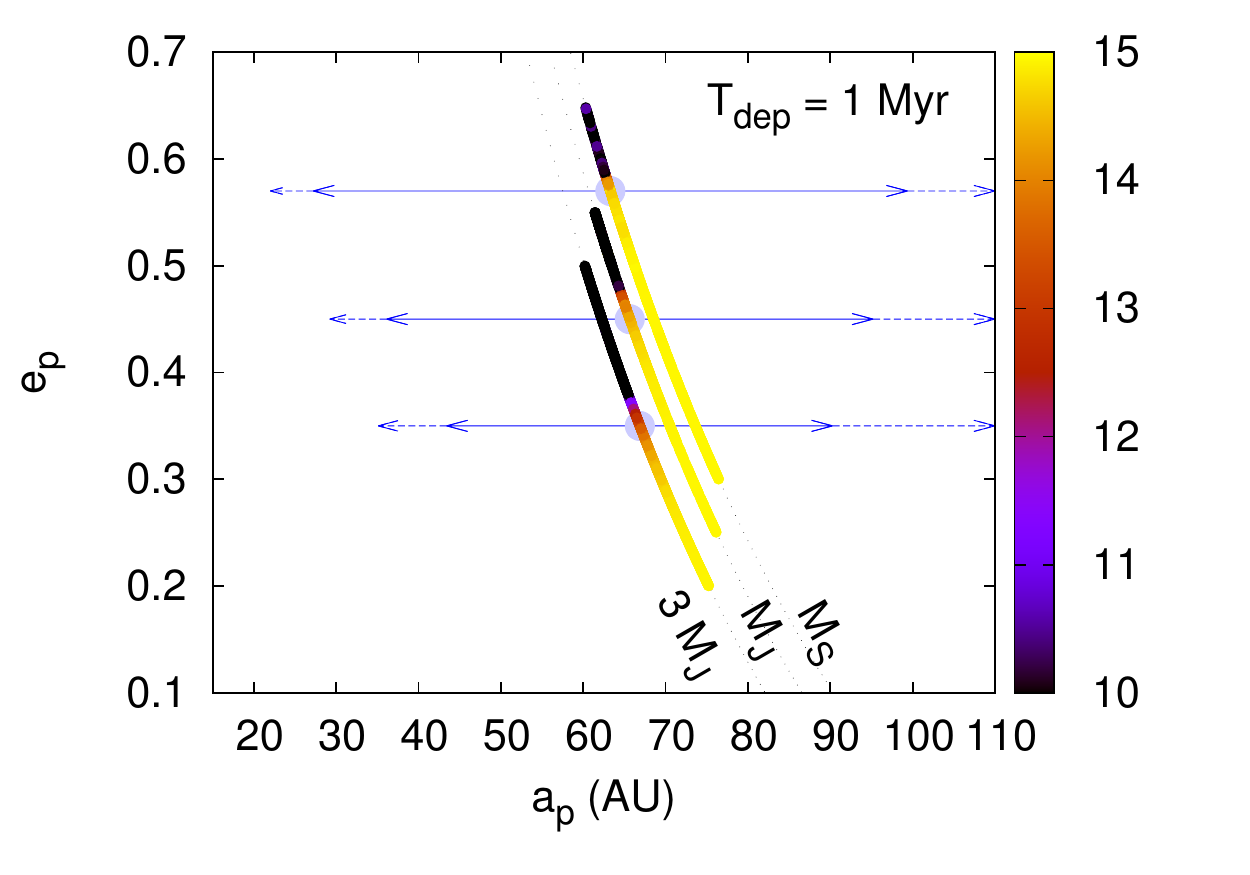} &
    \includegraphics[width=0.5\textwidth]{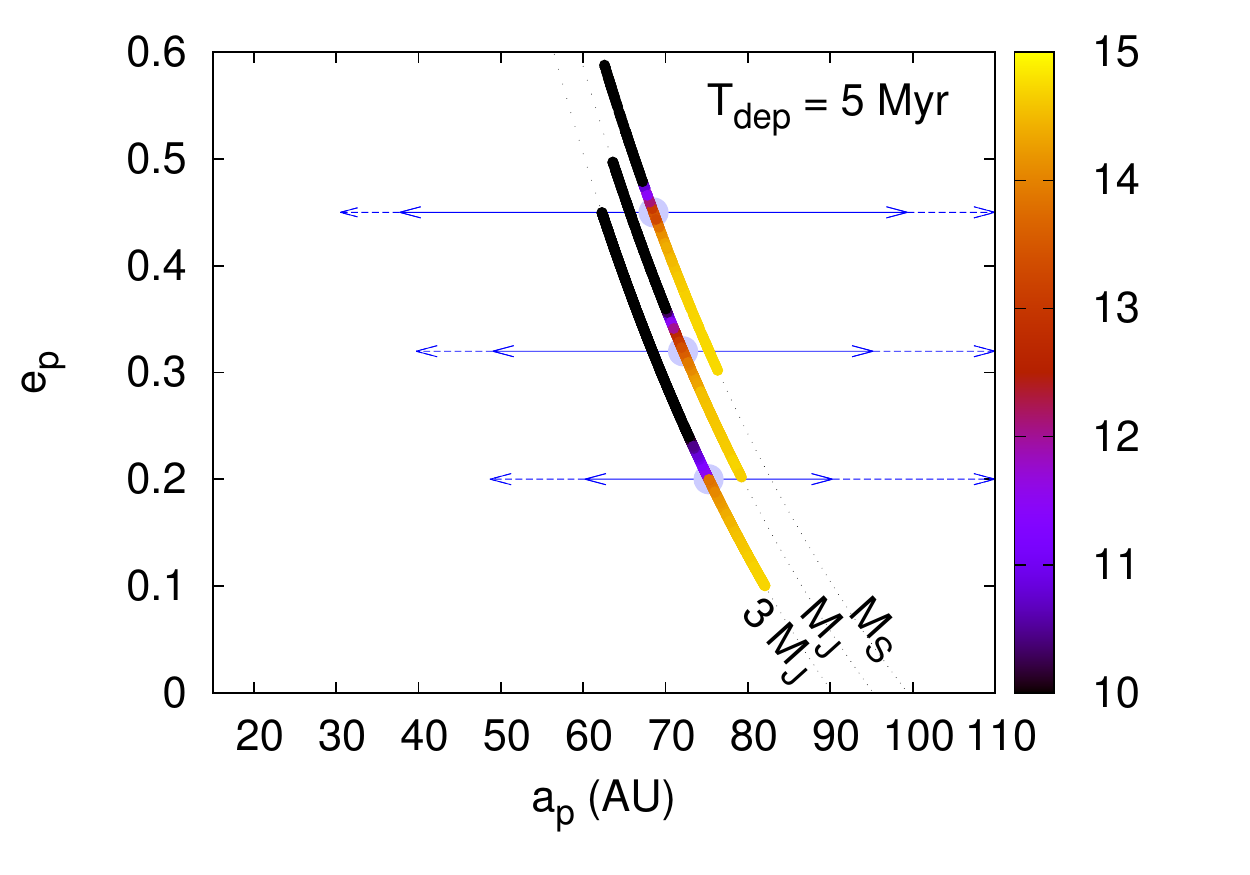}  \\
  \end{tabular}
\caption{The location of planetesimals after the passage of a hypothetical 
planet's sweeping secular resonance, as a function of its orbital parameters.
In order to test impact of the sweeping secular resonance on Vega's debris 
disk \citep{su2013}, a set of test planetesimals are initially placed at 
its inner boundary, i.e. 15 AU. Their locations at $t = 10 T_{\rm dep}$ 
are represented by colors with the scale indicated in the reference bar 
on the right hand side of the left (for $T_{\rm dep}=1$ Myr) and right (for 
$T_{\rm dep}=5$ Myr) panels. The light blue dots label the critical values 
of $a_{\rm p}$ and $e_{\rm p}$ which demarcate the planet's kinematic 
properties which may lead to significant orbital evolution for the perturbed
planetesimals. The solid blue line (with arrows) indicate the extent of 
an eccentric planet's radial excursion.  The dashed blue arrows refer its 
chaotic zone as calculated using Equation~(\ref{eq:a_ext}).}
\label{fig:aj_ej_tdep}
\end{figure*}

These considerations indicate that after the gas in a disk with an embedded
distant giant planet is severely depleted, a wide gap is expected to form 
between an inner warm and an outer cold belt.  The outer edge of the 
warm belt is mostly sculpted by the planet's sweeping secular resonance 
whereas the inner cavity of the cold belt essentially extends throughout
the planet's exterior chaotic zone. Based on this conjecture, we use the 
observed gap structure in the Vega system to place several quantitative 
constraints on the embedded planet's mass $m_{\rm p}$, semi-major axis $a_{\rm p}$ and 
eccentricity $e_{\rm p}$ (Fig.~\ref{fig:aj_ej_tdep}). 

We first satisfy the constraints set by the outer boundary of the detected 
dust-free gap (the inner boundary of the cold belt) $r_{\rm out} = 110$ AU
under the assumption $r_{\rm out} \simeq a_{\rm ext}$.  From Equation 
\ref{eq:a_ext}, we find an $a_{\rm p}-e_{\rm p}$ relation for each of 
three representative planet masses: $\msaturn$ (Saturn mass), $\mjupiter$ 
(Jupiter mass), and $3 \mjupiter$ which is an observational upper mass 
limit for any hypothetical planet in the region between $\sim 20$~AU 
and $\sim 70$~AU around Vega \citep{marois2006, heinze2008}. 

For each set of planetary orbital parameters, we find effects of the 
planet's secular resonance on the residual planetesimals interior to 
the planet's orbit. In general, the location and clearing efficiency of 
the secular resonance are determined by $\Sigma (r, t)$, $m_{\rm p}$, 
$a_{\rm p}$, and $e_{\rm p}$ \citep{nagasawa2003, nagasawa2005, 
thommes2008}. With a generic prescription for $\Sigma (r, t)$ (Eq. 
\ref{eq:sigma}) we compute the dynamical evolution of planetesimals'
orbits as the disk becomes severely depleted at $t = 10 T_{\rm dep}$.  
Two sets of $T_{\rm dep}$ (1 and 5 Myr) are used to evaluate whether 
the outcome may depend on its magnitude.  In all cases, the 
inner secular resonance has propagated well inside 10 AU with a 
local $\dampingtime < \depletiontime$ while the outer secular 
resonance has propagated well outside 150 AU with a local 
$\dampingtime > 10^3 \depletiontime$ at this epoch.

In order to examine whether the sweeping secular resonance can actually 
clear a region down to the observed inner boundary of the gap $r_{\rm in} = 15$ AU,
we consider a population of test planetesimals with an
initial circular orbit at 15 AU from the host star. There are three potential
outcomes for the perturbed planetesimals: (i) those that are excited 
by secular resonance and experience efficient damping, resulting in 
an inward migration; (ii) those that are insignificantly perturbed by
to the secular resonance and remain in the proximity of their original 
locations; and (iii) those that are highly excited by secular resonance 
but the gas drag is ineffective to damp their eccentricity, resulting in 
escape from the system. 

For the purpose of placing constraints on the hypothetical planet's 
orbital properties with the observed structure of Vega's debris disk, 
we neglect category (iii) and distinguish the retained planetesimals 
in categories (i) and (ii) by their final semi-major axis at $t = 10 
T_{\rm dep}$ (indicated by the color bar in Figure~\ref{fig:aj_ej_tdep}).  
These results indicate that the sweeping secular resonance of a lone 
planet with modest mass can effectively excite planetesimals' eccentricity 
and induce significant orbital decay provided it has adequate eccentricity.  
For each planet with a mass $m_{\rm p}$, there is a set of critical 
$a_{\rm p}$ and $e_{\rm p}$ (marked by light blue filled circles in 
Figure~\ref{fig:aj_ej_tdep}) which delineates categories (i) and (ii).
The critical values of $a_{\rm p}$ (in Fig.~\ref{fig:aj_ej_tdep}) 
is $\sim 63-67$ AU for all three different $m_{\rm p}$ with $T_{\rm dep} 
= 1$ Myr whereas those of $e_{\rm p}$ decreases from 0.57 for 
$m_{\rm p}= M_{\rm S}$ to 0.33 for $m_{\rm p}= 3 M_{\rm J}$.
With a longer $T_{\rm dep}~(= 5$ Myr), the critical values of 
$a_{\rm p} \simeq 68-75$ AU and $e_{\rm p} \simeq 0.45-0.2$ for
$m_{\rm p} = M_{\rm S}-3 M_{\rm J}$.  

\subsection{Single Planet Architecture}

\begin{figure*}
\centering
   \includegraphics[width=1\linewidth]{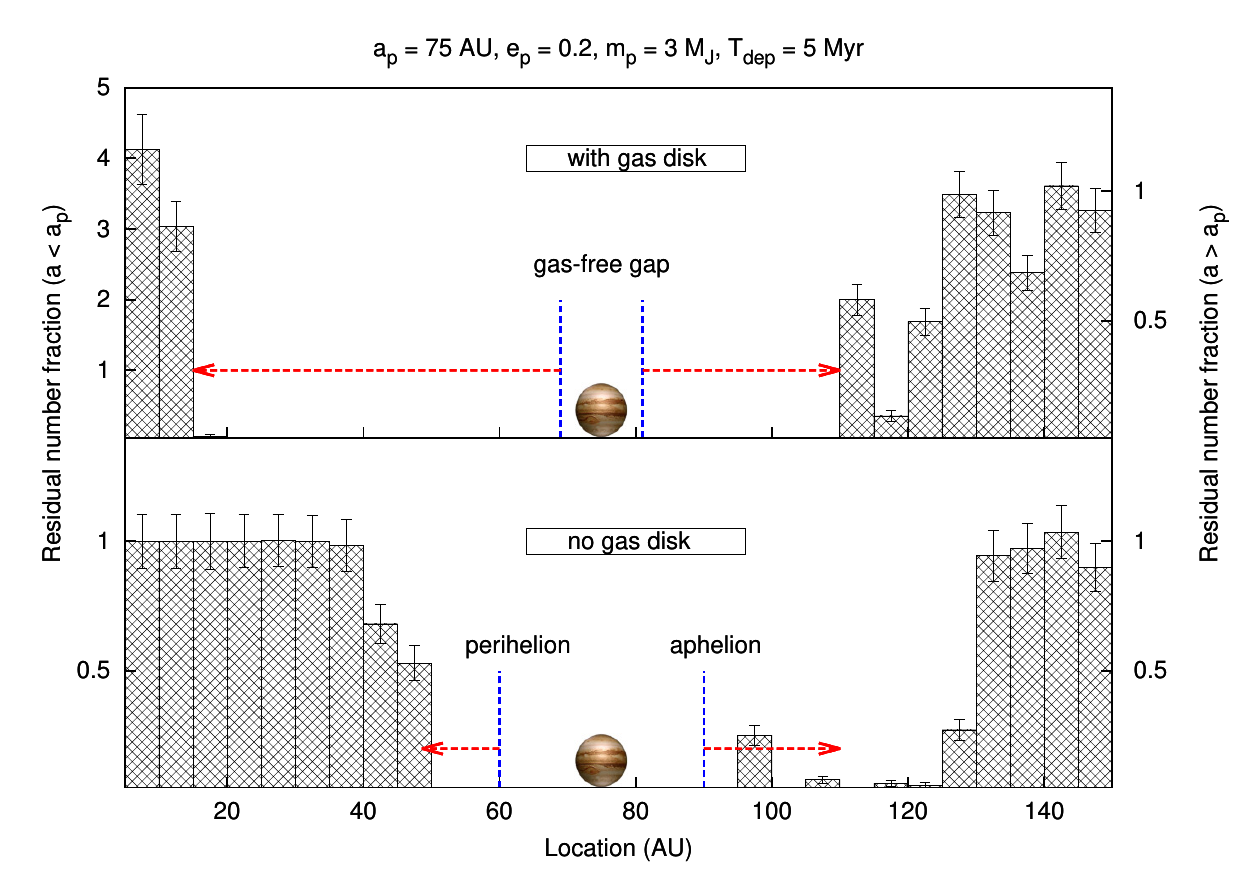} \\
\caption{Residual number fraction as a function of planetesimals' 
location. The top panel shows the model with gas disk, and includes 
the gravitational potential and gas drag effect, while the bottom 
panel shows the results of pure $N$-body interactions without a gas 
disk. The planet symbol indicates the semi-major axis of the giant planet 
(not to scale). In the top panel, the blue lines map out the gas-free zone around the 
planet, and red arrows map out the region where planetesimals
are expected to be completely cleared (dust-free gap boundary indicate by observation). In the bottom panel, blue lines in the no-gas model, label the perihelion and aphelion of the planet, and red arrows point to the boundary of the chaotic zone.}
\label{fig:a_df}
\end{figure*}

In the previous section, we placed several constraints
on the hypothetical planet's orbital parameters with a 
single planetesimal which was initially placed at 15 AU.  
In this section, we consider the evolution of 
planetesimals throughout the disk with a particular 
set of planetary orbital parameters.  We adopt the
observational upper limit ($3 \mjupiter$) for $M_p$.
Based on the results in Figure~\ref{fig:aj_ej_tdep}, we adopt 
$a_{\rm p} = 75$ AU, $e_{\rm p} = 0.2$, and $T_{\rm dep} = 5$ 
Myr. We place a population of $9 \times 10^3$ representative 
planetesimals with a uniform semi-major axis and azimuthal 
distribution between $5-150$~AU from the host star. These 
planetesimals all have zero initial eccentricity. Although 
the planet's inner and outer secular resonance sweep past 
$a_{\rm int}$ and $a_{\rm ext}$ within $t= 3 T_{\rm dep}$ (Fig. 
\ref{fig:ssrlocation}), we compute the planetesimals' orbital 
evolution to $t=10 T_{\rm dep}$ and plot, in Figure \ref{fig:a_df},
their retention fraction in equally-spaced (5 AU) bins of semi-major axis.  
This retention fraction is statistically computed 
from the ratio of final to initial number of planetesimals in 
each bin.

In order to distinguish between the effects of dynamical instability
and sweeping secular resonances, we carry out additional series of 
simulations of a purely $N$-body system.  Without the contribution to
the potential from any residual gas, the hypothetical planet does 
not precess and induce secular resonances to the planetesimals.  Nevertheless,
it induces both main motion and eccentric resonances to destabilize the
orbits of nearby planetesimals.  

In the absence of a gas potential, the survival fraction of planetesimals is mainly dominated by two competing gravitational interactions, that of the host star and that of the giant planet, respectively.  In Figure~\ref{fig:a_df}, the bottom panel shows that under the perturbation of gas giant, most planetesimals ($> 95 \%$)  within the planet's chaotic zone (red arrows label region separated from perihelion and aphelion) are scattered from their original locations. Most planetesimals can survive within 50~AU or external to 110~AU (two exceptions around 98~AU and 105~AU), roughly beyond the chaotic zone. It explains the observed features of cold belt's truncated region, but fails to fit the outer boundary of the warm belt in infrared observation ($\sim 15$~AU) under the lone-planet's scattering. However, taking a depleting gas disk into consideration, a large amount of planetesimals (especially for those within the orbit of giant planet) are excited along the sweeping path of secular resonance and subsequently orbital decay by gas drag, thus an extended planetesimal-free region is cleaned out, shown in the top panel of Figure~\ref{fig:a_df}. 
 
 As most of planetesimals within 45~AU (inner boundary of chaotic zone) are swept inward and assemble within 15~AU, this result is consistent with the observed boundaries of warm belt in Vega system. Also, within 15~AU, the number of planetesimals has roughly quadrupled by the end of the simulation. The latter suggests frequent collisions in this region, and perhaps that even the formation of super-Earths may be possible in the warm belt. \cite{thommes2008} studied the formation of the terrestrial planets in our Solar system resulting from the sweeping secular resonances of the giant planets, and suggests that a planet may form or is in the process of forming in the warm debris disk of Vega.

In the simulations with a circumstellar gas disk, the location of the outer mass collection is consistent with the observed infrared boundaries of the cold belt in Vega system \citep{su2013}. This indicates that even though the secular resonance sweeps through the gas disk in both directions (both inwards and outwards), the clearing effect caused by the sweeping secular resonance mechanism in the cold belt region is ineffective. It can be ignored because the inward migrations of planetesimals caused by damping is in the opposite direction to the secular resonance sweeping. Besides, the surface density of gas disk at large distances declines rapidly, making it hard to damp planetesimals and hence their orbital decay can be ignored. Therefore, applying the outer chaotic zone as one of boundary constraints to confine the orbital parameter of an unseen planet is reasonable and valid in the lone-planet scenario. 

\begin{figure*}
\centering
\includegraphics[width=1\textwidth,clip=true]{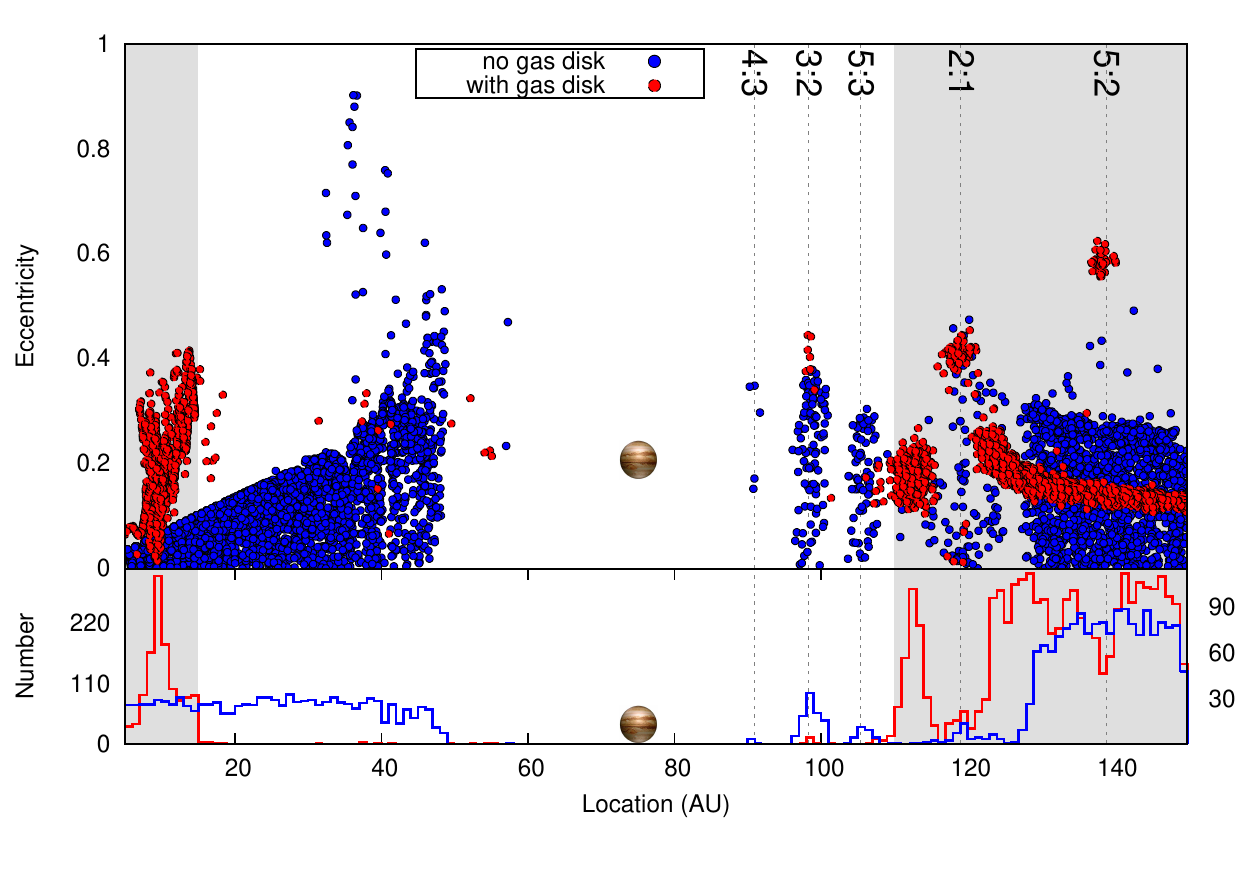}
\caption{The distribution of test planetesimals at time 50 Myr. A three Jupiter mass planet is located at 75 AU with eccentricity equal to $e=0.2$. The red and blue colors label the case that planetesimals evolve in a depleting gas disk and without a gas disk, respectively. The top panel shows the eccentricity distribution of planetesimals as a function of their locations, while the bottom panel statistically counts the planetesimals' number distribution in detail (planetesimals whose semi-major axis inside and outside of the giant planet are plotted separately, as indicated by the left and right axes).  The planet symbol indicates the semi-major axis of the giant planet (not to scale). Several mean motion resonance locations in the cold belt region are indicated with the dashed vertical lines, and the warm and cold debris belt regions are marked by the grey-shaded areas.}
\label{fig:ap_ep}
\end{figure*}

Interestingly, in pure $N$-body simulations, the mass assembly in the cold belt region is not completely truncated at 110~AU which is estimated to be chaotic boundary, there is still a non-negligible number of planetesimals isolated gathering at several semi-major axes between roughly 95 AU and 110 AU. To understand this anomaly in the mass assembly, we further discuss the distribution of planetesimals' orbital parameters after 50 Myr of evolution. In Figure~\ref{fig:ap_ep}, detailed accounting for residual planetesimals shows that even though a planet can scatter most planetesimals from its chaotic zone, in the gas-free environment, there are still some `lucky' planetesimals that can survive as they are captured into a powerful mean motion resonance, for example the 3:2 and 5:3 resonances, of the giant planet. Outside the chaotic zone, the eccentricities of planetesimals oscillate with large uncertainty due to the giant planet's secular perturbation. While in a depleting gas disk, planetesimals which are swept through and captured by the secular resonance of giant planet are mostly excited to some certain eccentricity and can maintain this value due to weak damping. It provides us a clue to make a rough judgement about whether the configuration of the two-belt debris disk is the result of sweeping secular resonance before the gas is completely depleted by measuring the eccentricity dispersion of the bodies in the cold belt.

\begin{figure*}
\centering
\includegraphics[width=1\textwidth,clip=true]{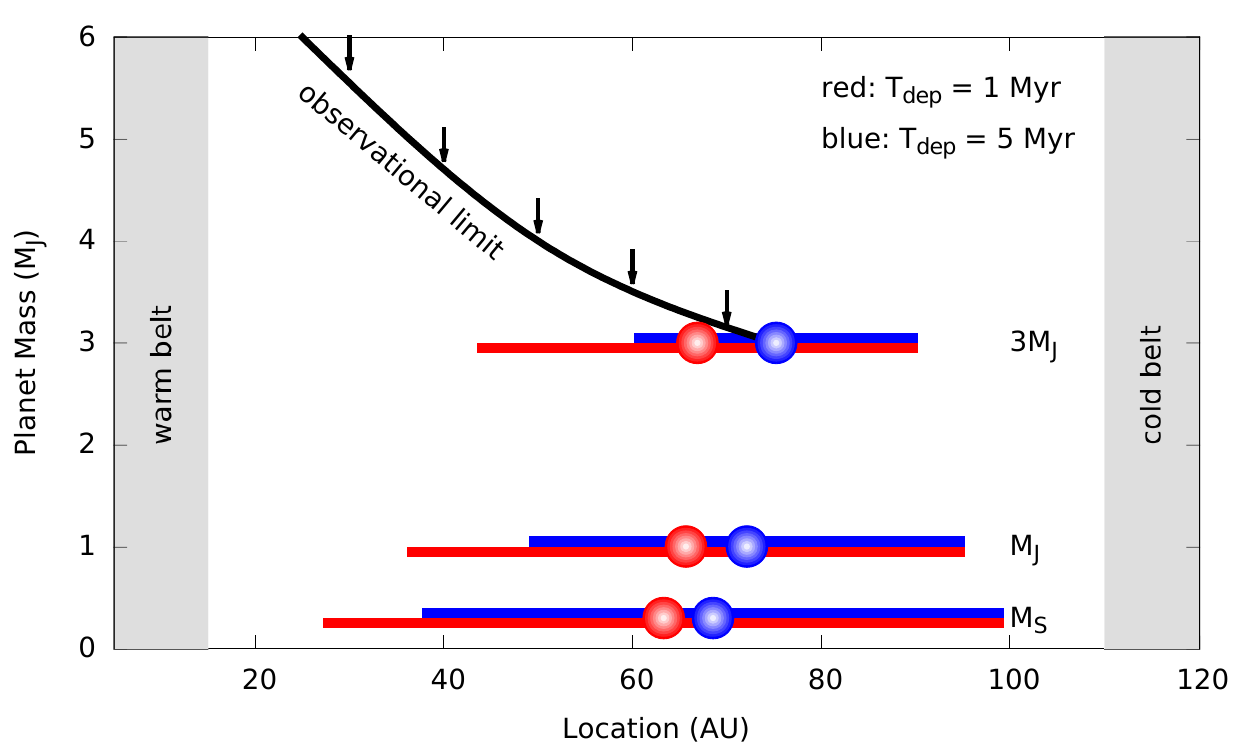}
\caption{A schematic view of the potential planetary architectures of the Vega system that can account for the large dust-free gap between warm and cold debris belts (grey shadowed regions). The red and blue dots present the possible semi-major axis and mass of an (unseen) planet according to the boundary constraints, while colored bars indicate the possible range of distances of each planet from the host star. Colors represent different evolving timescales of the gas disk, $T_{\rm dep} = 1$~Myr (red) and $T_{\rm dep} = 5$~Myr (blue). The observational limit from direct imaging for planets around Vega result from \cite{marois2006} and \cite{heinze2008} is also indicated. }
\label{fig:aj_mj}
\end{figure*}

In Figure~\ref{fig:aj_mj}, we illustrate the possible configurations which may account for the entire $\sim 15 - 110$~AU dust free gap in the Vega system. Considering the depletion timescale of the gas disk in the Vega system likely varies from $\sim 1$~Myr to $\sim 5$~ Myr (with a large uncertainty), the possible planet candidates are also expected to be detected in wide range of regions, especially for a low mass planet, e.g., a Saturn-mass planet which can generate observable features in our  scenario can be between $\sim 25$~AU and $\sim 100$~AU from the host star. Even for a $3 M_{\rm J}$ mass planet (upper limit), its detectable distance is $\sim 45 - 90$~AU.

%



\section{Application to the Fomalhaut system}
\label{sec:fomalhaut}

The Fomalhaut system is often treated as a sibling of the Vega system, since both systems have a host star of spectral type A, similar masses and ages, and both host a similar debris disk \citep{su2013}. According to \cite{su2013}, the dust-free gap in the debris disk of Fomalhaut extends from  $\sim 10$ AU to $\sim 140$ AU, which is slightly larger than the gap of the Vega system. It is therefore worthwhile considering a similar formation scenario for the dust-free gap in the Fomalhaut system.

\begin{figure*}
\centering
  \begin{tabular}{p{0.5\textwidth}p{0.5\textwidth}}
   \includegraphics[width=0.5\textwidth]{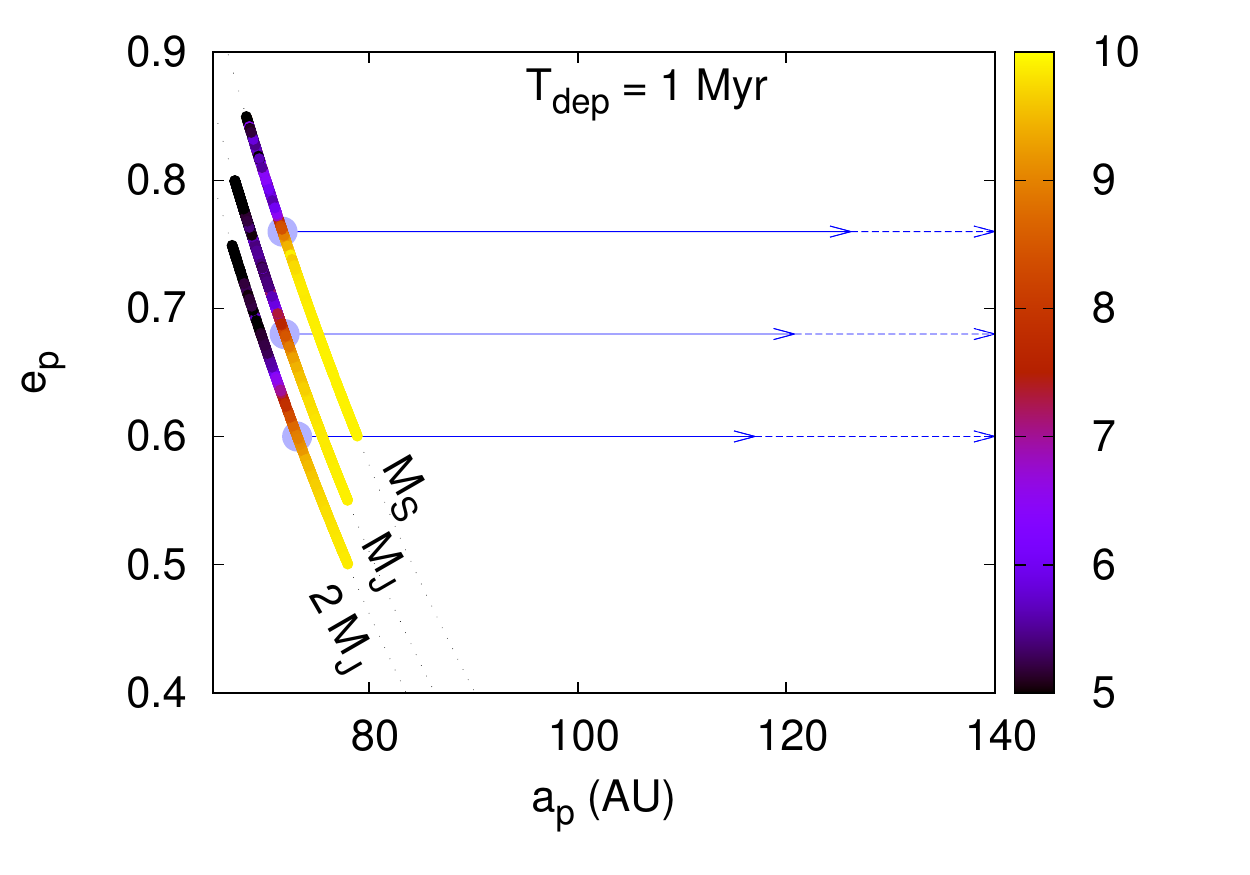} &
   \includegraphics[width=0.5\textwidth]{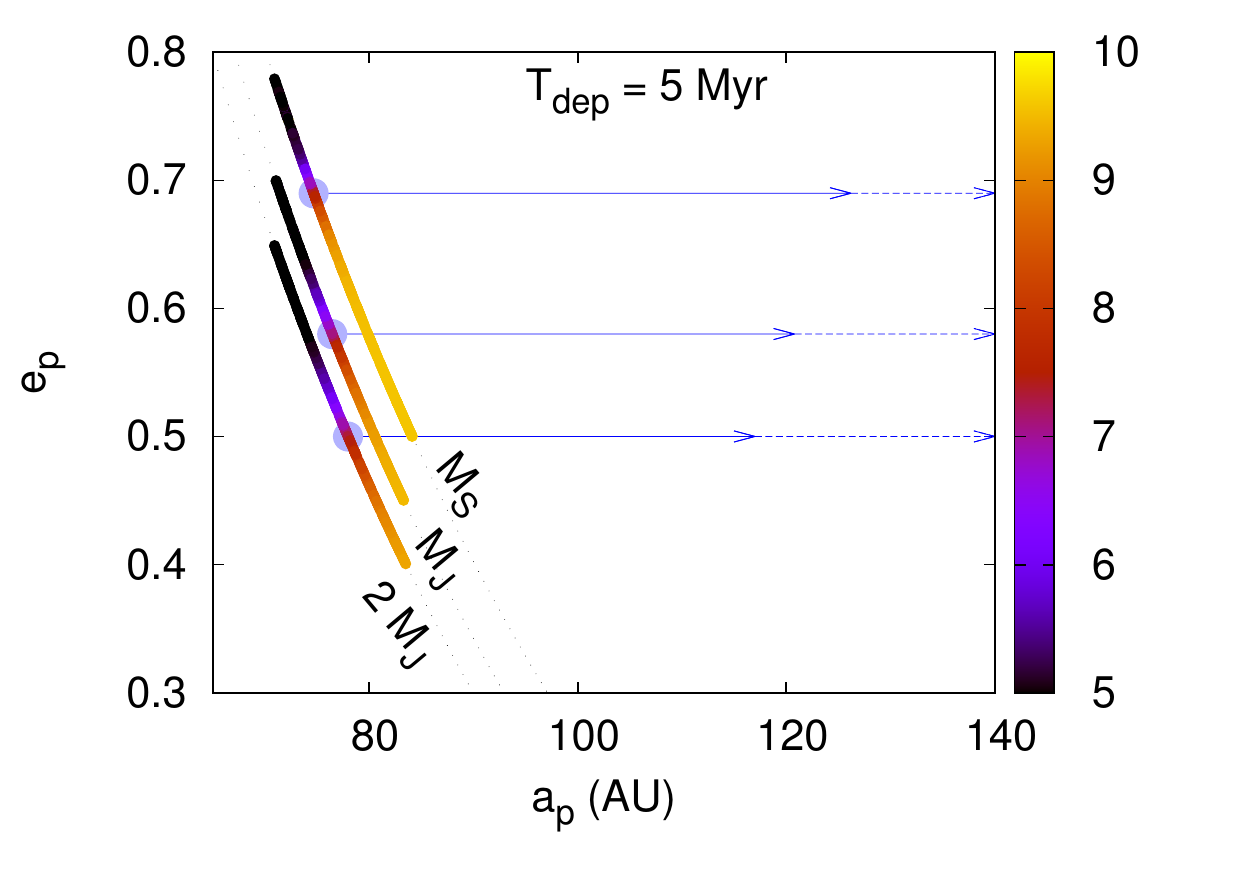}  \\
  \end{tabular}
\caption{As in Figure~\ref{fig:aj_ej_tdep}, a depleting gas disk with two dissipation timescales, 1~Myr and 5~Myr, are discussed separately.}
\label{fig:aj_ej_tdep_fomal}
\end{figure*}

Similar to the method applied to the Vega system, we primarily set 
some test planetesimals located at 10~AU to constrain the inner boundary. 
The outer boundary is based on the calculation using Equation~\ref{eq:a_ext}, 
assuming the apocenter of the gas giant is one chaotic zone width from 
the cold belt. And as inferred by \cite{kenworthy2009}, the ground-based 
high-contrast observations provide us a $2 \mjupiter$ upper mass limit 
for any planet located between $10 - 40$~AU. Therefore, we mainly 
explore the possibility of a hypothetical lone-planet with a mass 
$M_p = \msaturn$, $\mjupiter$, $2 \mjupiter$.  We vary the planet's 
orbital eccentricity $e_p$, semi-major axis $a_p$, and disk depletion
time scale $T_{\rm dep}$ in an attempt to reproduce the observed 
two-belt structure in Figure~\ref{fig:aj_ej_tdep_fomal}. In comparison
with the Vega system, our successful models require a large eccentricity 
and a relatively large ($\sim 70-80$ AU) semi-major axis to open 
the wide dust-free gap in the Fomalhaut system.

\begin{figure*}
\centering
\includegraphics[width=1\textwidth,clip=true]{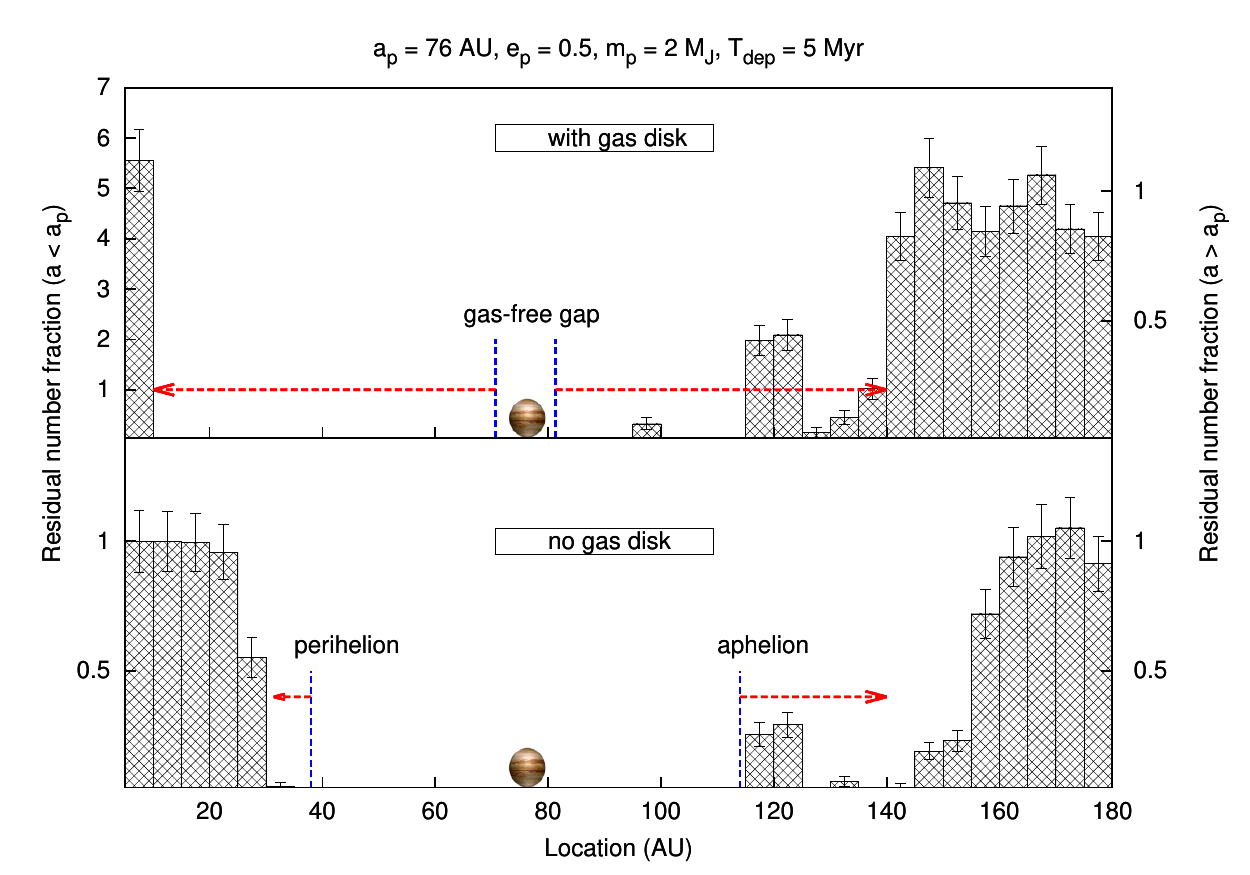}
\caption{As in Figure~\ref{fig:a_df}, but with a planet mass of two-Jupiter masses, which has a semi-major axis of 76 AU and an eccentricity of 0.5.}
\label{fig:a_df_fomal}
\end{figure*}

\begin{figure*}
\centering
\includegraphics[width=1\textwidth,clip=true]{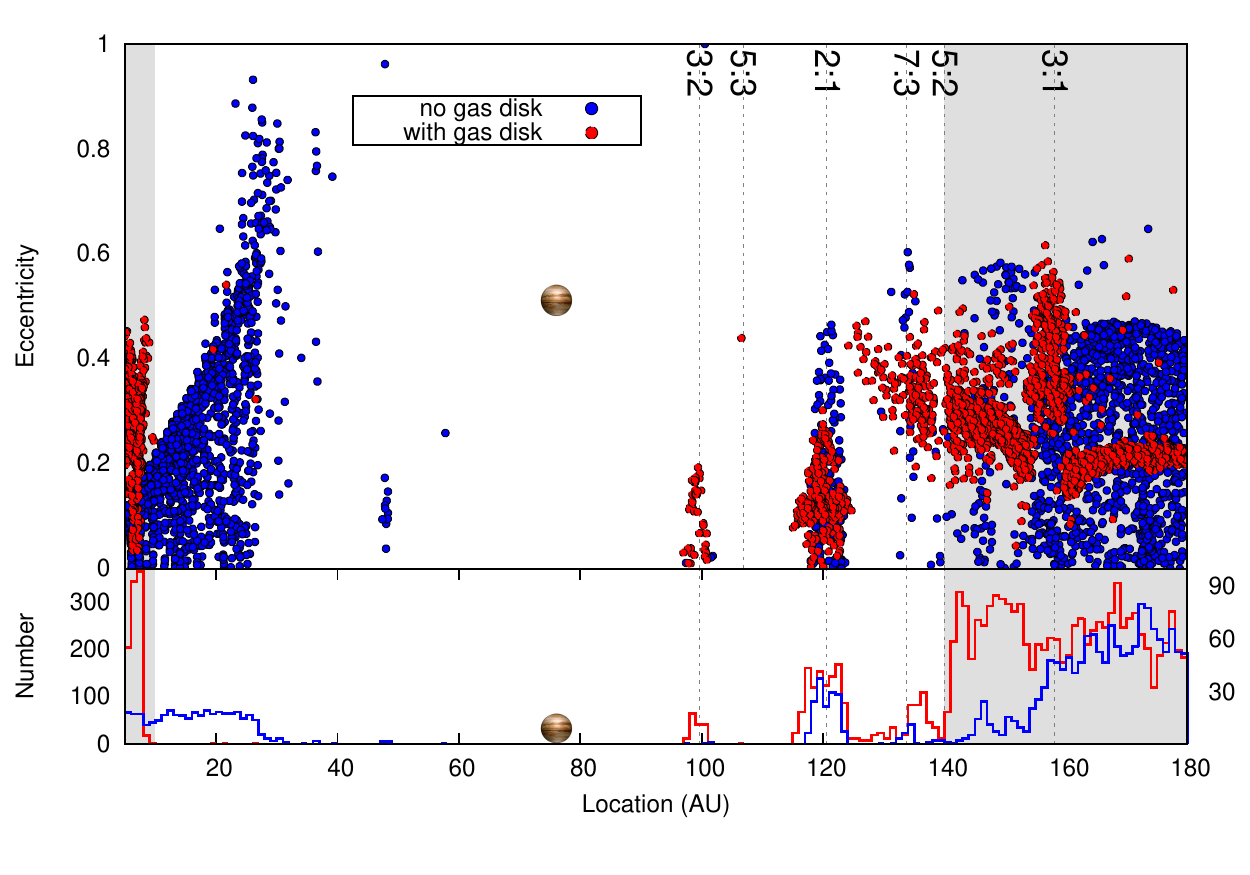}
\caption{As in Figure~\ref{fig:ap_ep}, but with a planet mass of two-Jupiter mass, which has a semi-major axis of 76 AU and an eccentricity of 0.5.}
\label{fig:ap_ep_fomal}
\end{figure*}

For example, a $M_p=2 \mjupiter$ planet can induce the observed gap
with $a_p \sim $76 AU and $e_p \sim 0.5$ in a disk with $T_{\rm dep} = 5$~Myr.
The results of other successful models are shown in 
Figures~\ref{fig:a_df_fomal} and \ref{fig:ap_ep_fomal}. These boundary 
conditions can lead to the severe clearing of planetesimals from the 
observed gap region and their migration to a region $\sim 10~$AU.
The population of residual planetesimals in the inner ring increases
by $\sim $5.5 times its original value. All planetesimals within
the planet's aphelion are cleared through orbit crossing and close 
encounters.  Beyond 140 AU, residual planetesimals are essentially
unperturbed and form an outer ring.  Near the inner boundary of the 
outer ring, some residual planetesimals accumulate near the planet's 
mean motion resonances, especially its 2:1 mean motion resonance. 
These local concentrations are separated from the continuous mass
distribution of the outer cold belt, and they may contribute to 
the subsequent formation of additional dwarf planets.

\begin{figure*}
\centering
\includegraphics[width=1\textwidth,clip=true]{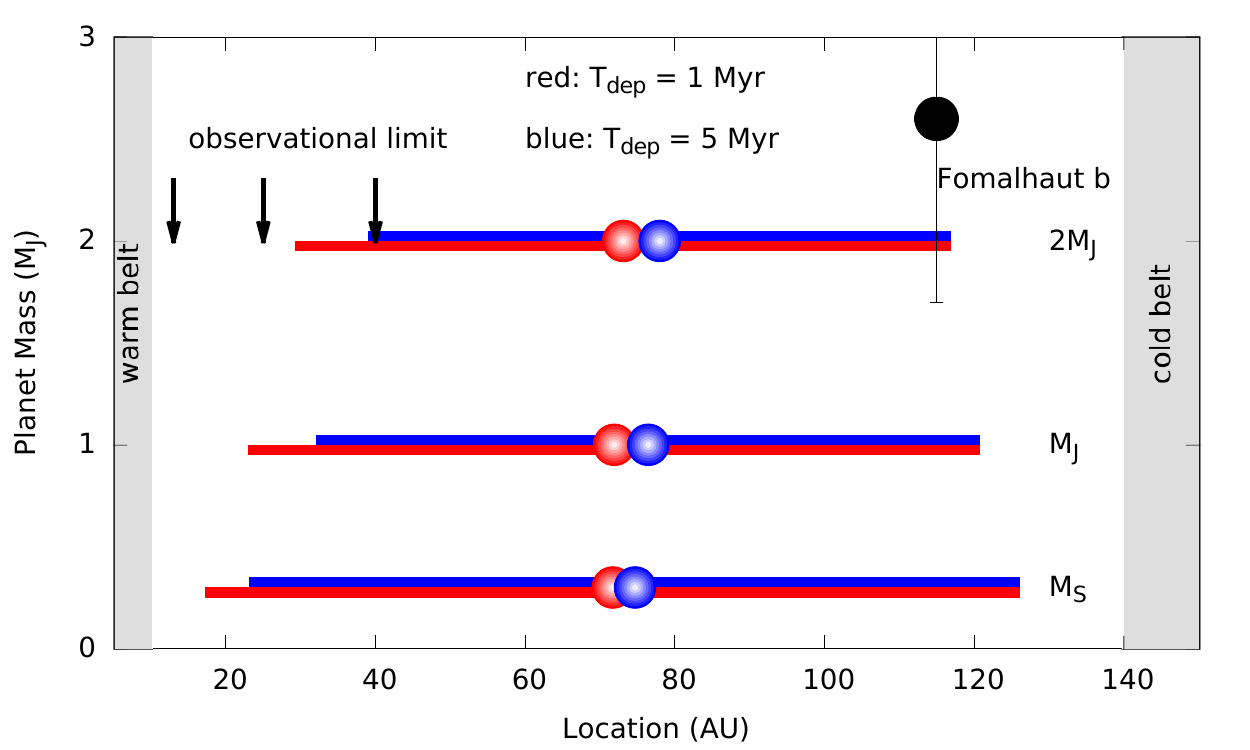}
\caption{An overall view of the potential planetary architecture in the Fomalhaut system, as Figure~\ref{fig:aj_mj}. According to \cite{kenworthy2009}, any planet orbiting between $\sim 15 -  40$ AU with mass $> 2 M_{\rm J}$ is detectable.}
\label{fig:aj_mj_fomal}
\end{figure*}

Our assumptions in the lone-planet scenario are not in conflict 
with what is known about the Fomalhaut system. According to the optical 
observations of \cite{kalas2008}, an exoplanet candidate was first 
presented in Fomalhaut system, named as Fomalhaut~b. This planet
may have dynamically sculpted the ring-like outer debris disk around 
Fomalhaut \citep{chiang2009}.  Even though the existence of this planet 
remains controversial \citep{lawler2015}, its observationally inferred
separation by \cite{kalas2008} ($\sim 115$~AU) is consistent with 
the estimated apastron of our hypothetical lone planet with 
$a_{\rm p} = 76~$AU and $e_{\rm p} = 0.5$ (see Figure~\ref{fig:aj_mj_fomal}). 
We note that the orbital data for Fomalhaut b in \emph{NASA Exoplanet 
Archive} suggest that the eccentricity of this candidate should be 
larger than 0.13 and its mass is $2.6 \pm 0.9$ times of the mass of Jupiter 
mass. Therefore, if the existence of Fomalhaut~b is indeed confirmed, 
it may be a good culprit for inducing the large dust-free gap in the 
debris disks of the Fomalhaut system.


\section{Summary and Discussions} \label{section:conclusions}

Infrared observations have indicated large dust-free gaps in the debris 
disks of both Vega and Fomalhaut. Up to now, the origin of these wide gaps
have been attributed to the dynamical perturbation of (yet to be detected) 
multiple gas giants. However, actual observational data do not support
this theoretical hypothesis.  With the exception of the HR8799 system, 
no Vega-like systems have been found to host multiple planets. In this 
paper, we put forward the lone-planet model.  We show that due to the sweeping
secular resonance effect, one planet with a modest eccentricity and mass
(a few $M_J$) is adequate to open very wide gap in the debris disk 
during the epoch of gas depletion.  

We apply the lone-planet hypothesis to the Vega system.  In this case, a 
$M_p= 3 M_J$ with a relatively small $e_p \sim 0.2$ and $a \sim 75$ AU would 
be sufficient to open and maintain a wide gap which resembles  
the observed debris distribution.  We constructed a similar model for 
the Fomalhaut system.  For the wider gap in this system, a $M_p=2M_J$ planet
with an eccentricity $e_p \sim 0.5$ and semi-major axis $a_p \sim 76$~AU, 
would successfully clear most of residual planetesimals from 10~AU to 140~AU
during the disk depletion. From these results, we infer that (1) the 
inner boundary of the cold belt is mainly cleared by the gas giant's 
gravity, from the cold belt near one chaotic zone width; (2) the outer 
boundary of the warm belt is dominated by secular resonance's sweeping 
path.

Finally, our results indicate that the powerful cleaning effect of 
the sweeping secular resonance along its path, does not only impact on the
kinematic configuration in our own Solar system but also influences
other exoplanet systems. All planetary systems are born in a circumstellar 
disk. A significant fraction ($\sim 15-20\%$) of solar type stars 
and larger fraction of more massive stars contain one or more gas giants. 
Up on the emergence of these massive planets, the sweeping secular resonance 
would occur naturally during the depletion of the disk provided they have
at least a modest eccentricity. As it propagates throughout the disk, 
eccentricity of a population of residual planetesimals is excited.  The depleting disk
gas also damp their eccentricity and induce their orbital decay.  The combined
influence of the sweeping secular resonance and gas damping of planetesimals can produce a large dust-free gap. In the current work, we used this lone-planet hypothesis to infer the origin 
of wide gaps in Vega and Fomalhaut systems, and it is also possible to extend 
this scenario to other double-ring debris disks in the future.


\section*{Acknowledgments}

{This work was supported by the National Science Foundation of China (Grant No. 11333003, 11390372).
M.B.N.K. was supported by the National Natural Science Foundation of China (grants 11010237, 11050110414, 11173004, and 11573004) and the Research Development Fund (grant RDF-16-01-16) of Xi'an Jiaotong-Liverpool University (XJTLU).}


\software{HERMIT4 \citep{aarseth2003}}



\end{CJK*}

\end{document}